\documentclass[aps,showpacs,amsmath,amssymb,prb,preprint]{revtex4-1}
\usepackage{color}
\usepackage{graphicx}% Include figure files
\usepackage{dcolumn}% Align table columns on decimal point
\usepackage{bm}% bold math
\usepackage[nolist,nohyperlinks]{acronym}
\usepackage{lmodern}
\usepackage{epstopdf}
\usepackage{subfigure}
\usepackage{amsmath} %math
\usepackage{braket}
\usepackage{natbib}
\newcommand{\CYS} {{\mbox{CdYb${}_2$Se${}_4$}}}
\newcommand{\TYX} {{\mbox{TYb${}_2$X${}_4$}}}
\newcommand{\ETO} {{\mbox{Er${}_2$Ti${}_2$O${}_7$}}}
\newcommand{\EGO} {{\mbox{Er${}_2$Ge${}_2$O${}_7$}}}
\newcommand{\YGO} {{\mbox{Yb${}_2$Ge${}_2$O${}_7$}}}
\newcommand{\YTO} {{\mbox{Yb${}_2$Ti${}_2$O${}_7$}}}

\newcommand{\half}{{\ensuremath{\frac{1}{2}}}}

\newcommand{\quat}{{\ensuremath{\frac{1}{4}}}}

\DeclareMathAlphabet{\mathpzc}{OT1}{pzc}{m}{it}

\newcommand{\avg}[1]{\braket{#1}}

\newcommand{\cc}[1]{{#1}^{*}}

\renewcommand{\vec}[1]{\boldsymbol{#1}}
\newcommand{\mat}[1]{\vec{#1}}
\newcommand{\trp}[1]{{#1}^{\intercal}}
\newcommand{\vhat}[1]{\vec{\hat{#1}}}

\newcommand{\meV}{\ {\rm meV}}
\newcommand{\K}{\ {\rm K}}

\begin{document}
%\preprint{APS}

\title{Multi-phase competition in quantum $XY$ pyrochlore antiferromagnet CdYb$_{2}$Se$_{4}$: zero and applied magnetic field study}

%%%%% BEGINNING OF AUTHOR LIST %%%%%
\author{K. Guratinder}
\affiliation{Laboratory for Neutron Scattering and Imaging, Paul Scherrer Institute, CH-5232 Villigen, Switzerland}
\affiliation{Department of Quantum Matter Physics, University of Geneva, CH-1211 Geneva, Switzerland}
\author{Jeffrey G. Rau}
\affiliation{Max-Planck-Institut f\"ur Physik komplexer Systeme, 01187 Dresden, Germany}
%\author{M. Pregelj}
%\affiliation{Institute "Jozef Stefan", Jamova 39, 1000 Ljubljana, Slovenia}
\author{V. Tsurkan}
\affiliation{Experimental Physics V, Center for Electronic Correlations and Magnetism, University of Augsburg, D-86159 Augsburg, Germany}
\affiliation{Institute of Applied Physics, Academy of Sciences of Moldova, MD-2028 Chisinau, Republic of Moldova}
\author{C. Ritter}
\affiliation{Institut Laue-Langevin, 156X, 38042 Grenoble C\'{e}dex, France}
\author{J. Embs}
\affiliation{Laboratory for Neutron Scattering and Imaging, Paul Scherrer Institute, CH-5232 Villigen, Switzerland }
\author{H. C. Walker}
\affiliation{Rutherford Appleton Laboratory, ISIS Facility, Chilton, Didcot, Oxon OX11 0QX, United Kingdom}
%\author{L. Keller}
%\affiliation{Laboratory for Neutron Scattering and Imaging, Paul Scherrer Institute, CH-5232 Villigen, Switzerland}
\author{M. Medarde}
\affiliation{Laboratory for Multiscale Materials Experiments, Paul Scherrer Institute, CH-5232 Villigen, Switzerland}
\author{T. Shang}
\affiliation{Laboratory for Multiscale Materials Experiments, Paul Scherrer Institute, CH-5232 Villigen, Switzerland}
\author{A. Cervellino}
\affiliation{Swiss Light Source, Paul Scherrer Institute, CH-5232 Villigen, Switzerland}
\author{T. Fennell}
\affiliation{Laboratory for Neutron Scattering and Imaging, Paul Scherrer Institute, CH-5232 Villigen, Switzerland }
\author{Ch. R\"{u}egg}
\affiliation{Research Division Neutrons and Muons, Paul Scherrer Institute, CH-5232 Villigen, Switzerland}
\affiliation{Department of Quantum Matter Physics, University of Geneva, CH-1211 Geneva, Switzerland}
\author{O. Zaharko}\email{oksana.zaharko@psi.ch}
\affiliation{Laboratory for Neutron Scattering and Imaging, Paul Scherrer Institute, CH-5232 Villigen, Switzerland.}
\date{\today}

\begin{abstract}
We study magnetic behaviour of the Yb$^{3+}$ ions on a frustrated pyrochlore lattice in the spinel {\CYS}.
The crystal-electric field parameters deduced from high-energy inelastic neutron scattering reveal
well-isolated ytterbium ground state doublet with a weakly Ising character.
Magnetic order studied by powder neutron diffraction evolves from the $XY$-type
antiferromagnetic $\Gamma_5$ state to a splayed ice-like ferromagnet (both with k=0)
in applied magnetic field with $B_c$=3 T. 
Low-energy inelastic neutron scattering identifies weakly dispersive magnetic bands around 0.72 meV 
starting at $\mid\bf{Q}\mid$ = 1.1 \AA$^{-1}$~ at zero field, which diminish with field and
vanish above 3 T.
We explain the observed magnetic behaviour in framework of the nearest-neighbour anisotropic 
exchange model for effective $S=1/2$ Kramers doublets on the pyrochlore lattice.
The estimated exchanges position the {\CYS} spinel close to the phase boundary between the $\Gamma_5$ 
and splayed ferromagnet states, similar to the Yb-pyrochlores suggesting an important role of the competition between these phases.
\end{abstract}

%\pacs{}
\maketitle
\section{Introduction}{\label{Sec1}}
The study of the family of rare-earth pyrochlores~\cite{Gardner2010} of the form R$_2$B$_2$O$_7$, with R a trivalent rare-earth and B a non-magnetic transition metal, has unveiled a rich variety of correlated magnetic states with unconventional excitations.
Examples include classical spin ices~\cite{GingrasSI} with emergent magnetic monopoles in Ho$_2$Ti$_2$O$_7$ and Dy$_2$Ti$_2$O$_7$, the physics of order-by-disorder in Er$_2$Ti$_2$O$_7$,~\cite{HallasARCMP,Rau2019} an unusually broad continuum of excitations in the anisotropic ferromagnet Yb$_2$Ti$_2$O$_7$,~\cite{HallasARCMP,Rau2019} and potential quantum spin liquid candidates in Tb$_2$Ti$_2$O$_7$~\cite{Rau2019}, Pr$_2$Hf$_2$O$_7$~\cite{Sibille2018} and Ce$_2$Zr$_2$O$_7$~\cite{CeZr1}.

Many of these materials are well described by an effective spin-1/2 model with anisotropic exchange interactions.
This effective spin-1/2 arises from the presence of a large crystal-electric field (CEF) splitting, isolating the ground doublet from the higher lying excitations~\cite{RepProgPhys2014}.
The low energy physics of these effective spin-1/2 models can exhibit strong quantum effects; this is especially true for trivalent ytterbium compounds, where the (relatively) low total angular momentum ($J=7/2$) allows for super-exchange interactions to induce quantum interactions, independent of the composition of the crystal field ground doublet~\cite{Rau2018}.

Interestingly, the same magnetic frustrated pyrochlore structure found in the R$_2$B$_2$O$_7$ compounds also arises in the \emph{spinels} with the composition TR$_2$X$_4$ (T=Cd, Mg, X=S, Se).
However, unlike the usual pyrochlores where the local environment is a distorted cube of eight oxygens, in the spinels the rare-earth ion is surrounded by a near perfect octahedron of X ligands~\cite{Lau2005}.
This structural difference distinguishes the spinels from the pyrochlores and allows for drastically different crystal field and exchange physics to arise, even for the same rare-earth element.
As an example, while the Er-pyrochlores are XY magnets~\cite{Champion2003}, the Er-spinels exhibit dipolar spin ice physics~\cite{Lago2010, Gao2018}.
Consequently, this family of materials has attracted attention as providing new potential routes to explore some of the rich physics seen in the rare-earth pyrochlores in a new setting.

The ytterbium spinels, of the form {\TYX}, in particular have attracted attention as possibly realizing highly frustrated quantum pyrochlore antiferromagnets~\cite{Higo2017, Reotier2017, Rau2018}.
Earlier experimental analysis of {\TYX}~\cite{Higo2017} found that all of the compounds order antiferromagnetically at $T_N \sim 1.6$ K -- much lower than the estimated exchange scale of $\sim 10$ K, suggesting strong frustration.
From the size of the local magnetic fields inferred from $\mu$SR experiments, only a small static moment of $\sim$ 0.1 $\mu_B$ was deduced.
Antiferromagnetic order was directly observed shortly afterwards~\cite{Reotier2017}, with neutron diffraction detecting magnetic Bragg peaks with propagation vector $\bf{k}$=0, corresponding to a magnetic state in the $\Gamma_5$ irreducible representation ~\cite{note0} with an ordered moment of $\sim  0.7 \mu_B$.
Together, these neutron and $\mu$SR results suggest that the magnetic state has a dynamic character.~\cite{Reotier2017}

In this article we present further experimental and theoretical exploration of {\CYS}. We obtain more detailed information about the CEF using inelastic neutron scattering in combination with magnetization and susceptibility measurements.
We then study how the magnetic order and spin dynamics evolve under the application of a magnetic field.
We compare our findings with expectations for an effective spin-1/2 anisotropic exchange model on the pyrochlore lattice, motivated by recent results on the super-exchange mechanism in related breathing pyrochlore materials~\cite{Rau2016b,Rau2018}.
From these considerations, we suggest that {\CYS} may share some physics with the ytterbium pyrochlores Yb$_2$M$_2$O$_7$ (M=Ti, Sn, Ge), including proximity to a phase boundary between antiferromagnetic and ferromagnetic phases.

\section{Experimental}{\label{Sec2}}
A polycrystalline sample of 4.75 g of $^{114}$CdYb$_{2}$Se$_{4}$ was prepared by solid state synthesis from binary Yb and Cd selenides. The $^{114}$Cd-isotope was used to reduce absorption in the neutron scattering experiments.
The purity and crystal structure of the sample were checked by synchrotron x-ray powder diffraction on
the Material Science beamline of Swiss Light Source SLS. The pattern collected with $\lambda$=0.565475 \AA~ at room temperature is well fitted with the normal spinel structure with no site inversion. 
Tiny amounts of Yb$_2$O$_2$Se and CdSe impurities (both below 1\%) were detected.

Magnetic susceptibility and magnetization were measured in the temperature ranges 0.5 K - 4 K and 1.8 K - 400 K in applied magnetic fields up to 7 T using a Magnetic Properties Measurement System superconducting quantum interference device (SQUID) magnetometer and a Physical Properties Measurement System, both from Quantum Design.
% Continuous wave electron-spin-resonance (ESR) measurements were performed on a 20 mg sample from room temperature down to 5\,K in the X-band (frequency $\sim$9.4 GHz) with a custom-built spectrometer equipped with a Varian E-101 microwave bridge.

High-energy time-of-flight inelastic neutron scattering (INS) measurements were performed on the MERLIN spectrometer\cite{Bewley2009} at ISIS. The powder sample, contained in an aluminium can in an annular geometry, was inserted into a closed cycle refrigerator. An incident neutron energy of $E_i$=150 meV and chopper frequency 450 Hz were used on cooling and then a setup with $E_i$=50 meV and chopper frequency of 500 Hz was chosen to record data at 8 K and 295 K. We used the McPhase program\cite{Rotter2004} to extract the crystal electric field parameters from the measured spectra. %and then verified the correctness of the model by a least-square fit of the CEF coefficients versus combined experimental information from bulk measurements and INS.

Neutron powder diffraction (NPD) experiments in zero and applied magnetic fields were performed on the DMC diffractometer at SINQ and on the high-flux diffractometer D20 at ILL ($\lambda$=2.41 \AA). 
We present here the results from D20. The polycrystalline sample was mixed with deuterated ethanol and methanol solution and placed in a double-wall Cu cylinder with inner and outer diameters 8 and 10 mm. The sample was then mounted  in an $^3$He insert in a 6 T magnet and cooled down to 0.45 K. Special care was taken to have zero applied field for the first cooling of the sample by demagnetizing the magnet before the usage. The Fullprof suite\cite{fullprof1993} was used to refine difference diffraction patterns between an ordered state at a given field $B$ at 0.45 K and the paramagnetic state at $T$=2.5 K measured in zero field.

Low-energy inelastic neutron scattering measurements were performed on the FOCUS spectrometer at SINQ. A double-wall Cu-can with the sample was inserted in a dilution refrigerator and a 10 T magnet. The base temperature of  0.1 K was reached. Two instrumental setups were used - with $E_i$= 10.4 meV (with elastic resolution $\delta E_0\approx$ 0.8 meV) and with $E_i$= 3.7 meV  ($\delta E_0\approx$ 0.35 meV). Empty can and vanadium corrections were performed.

\section{Results}{\label{Sec3}}
\subsection{INS measurement of CEF excitations}{\label{Sec3Sub1}}

\begin{figure}
\centering 
\includegraphics[width=0.95\columnwidth]{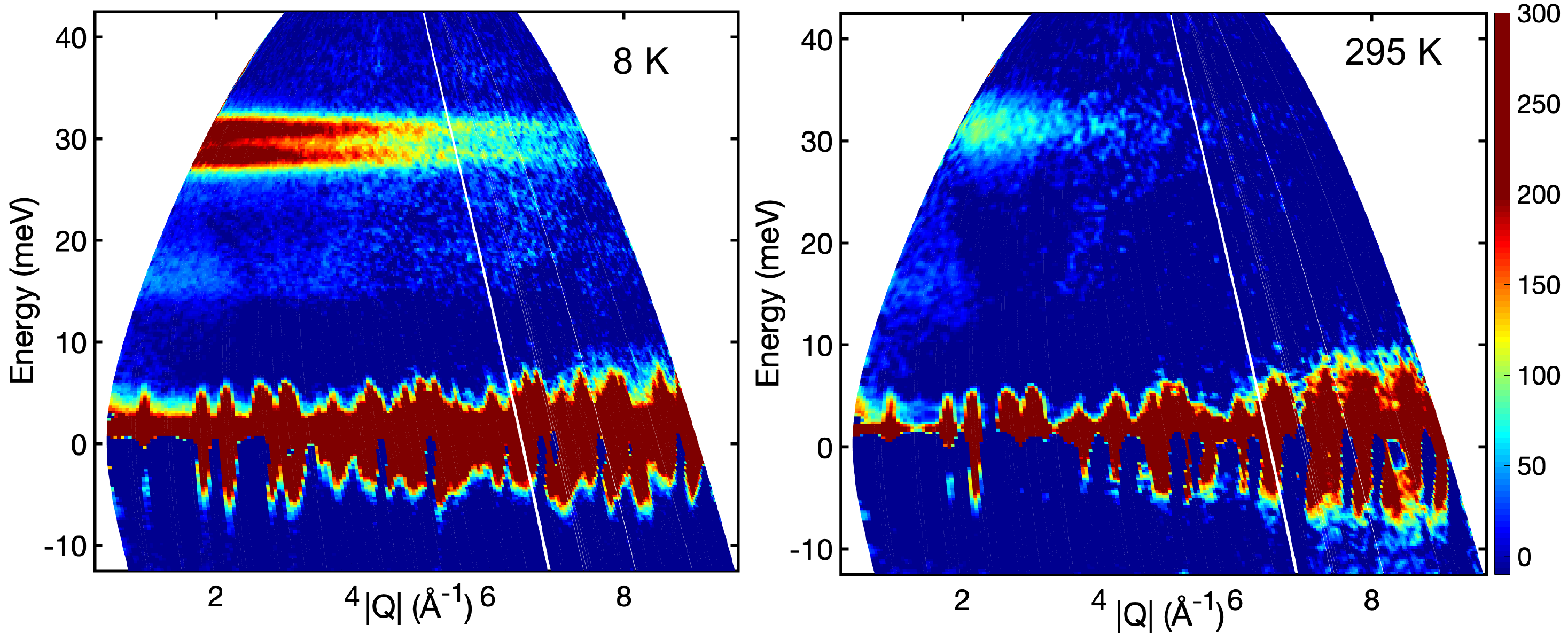} 
\caption{High-energy inelastic neutron scattering spectra at  8 K (left) and 295 K (right) measured on the MERLIN spectrometer with $E_i$= 50 meV.}
\label{fig2} 
\end{figure}

The INS spectra measured on MERLIN ($E_i$=50 meV) contain two dispersionless excitations at 28.33(2) meV and 31.07(2) meV, which have highest intensity at low momentum transfer $\mid\bf{Q}\mid$ and at low temperature (Fig.~\ref{fig2}). Therefore we attribute them to CEF transitions. These observations agree with the four doublet level scheme expected for Yb$^{3+}$-ion in chalcogenide spinels.\cite{Higo2017, Rau2018} The ground state (GS) doublet is separated by a significant gap of $\sim$ 30 meV from two of the three higher lying doublets.

The excitation at 18.8(7) meV is pronounced at high Q and at high temperature.
By comparison with measurements of an empty aluminium can (not shown here) we attribute this signal to aluminium phonons.

We did not observe any high-energy excitations with the $E_i$=150 meV setup, suggesting that the transition to the highest lying CEF level
has very small intensity, below the experimental sensitivity.

\subsection{Bulk measurements}
\begin{figure}
\centering 
\includegraphics[width=0.6\columnwidth]{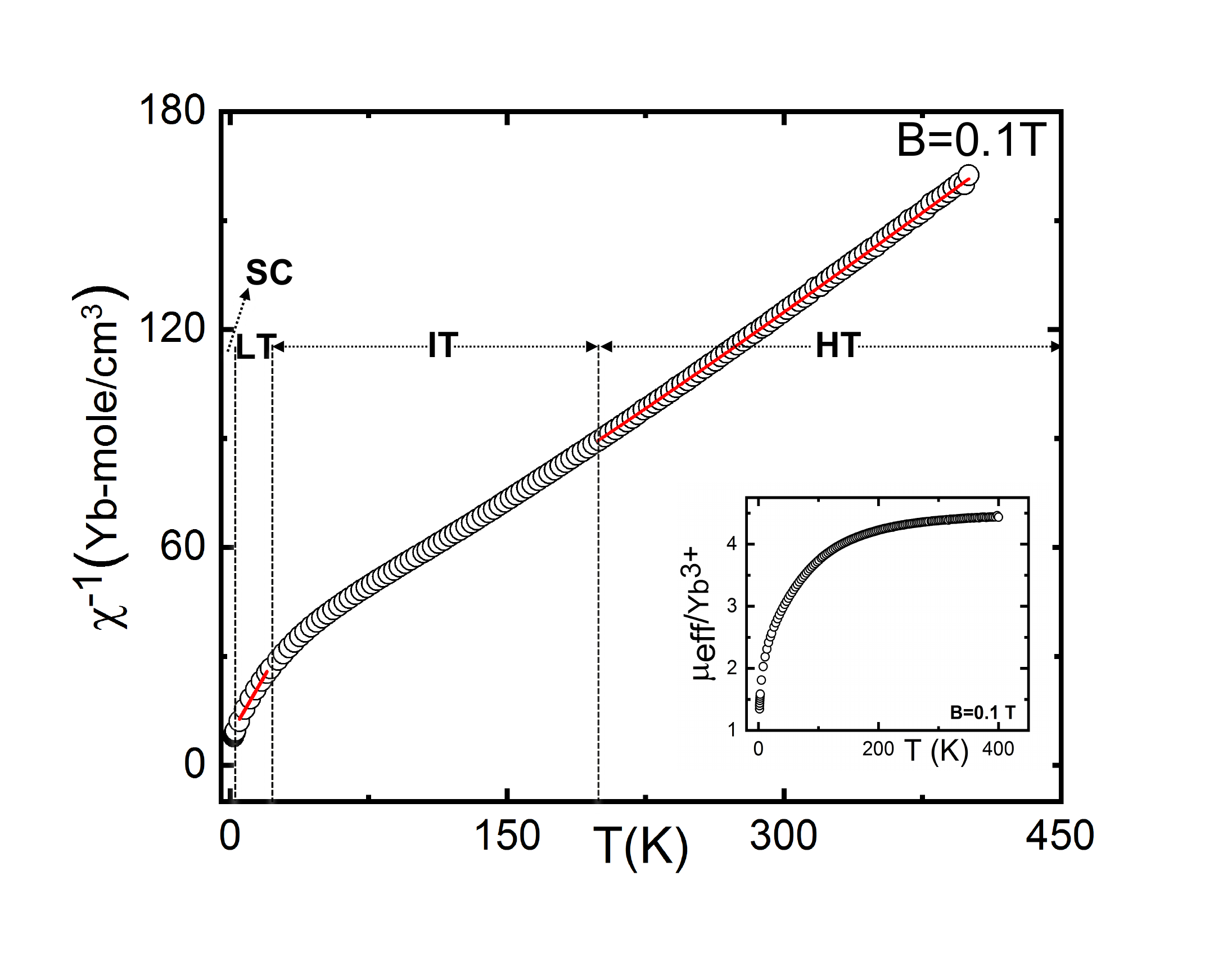} 
\caption{Inverse susceptibility ${1/\chi}$ in the four regimes measured at B=0.1 T and (insert) effective moment $\mu_{eff}$ versus temperature. The red solid lines are the linear Curie-Weiss fits.
}
\label{fig7} 
\end{figure}

{\label{Sec3Sub2}}
The inverse magnetic susceptibility, $\chi^{-1}$, of our polycrystalline sample (Fig.~\ref{fig7}) is in accordance with the published results.~\cite{Reotier2017, Higo2017} The temperature dependence $\chi^{-1}$ can be divided into four regions:
the 200 K $<T<$ 400 K high-temperature (HT) linear regime, 20 K $<T<$ 200 K intermediate-temperature (IT) non-linear regime, 5 K $<T<$  20 K  low-temperature (LT) linear correlated paramagnet regime and the strongly correlated (SC) regime  $T<  5$ K.

The HT linear regime is dominated by the susceptibility of the Yb$^{3+}$ ions in their ground state, because the contribution of the two next excited states does not exceed $10\%$. We use this HT regime to corroborate the set of the CEF parameters deduced from INS.

The IT regime is not linear and the $\chi^{-1}$ slope changes significantly near 70 K, probably due to the development of spin correlations. Here it is complex to disentangle the CEF and exchange contributions to the susceptibility.

The LT linear regime is governed by exchange correlations between the Yb$^{3+}$ ions with the $S=1/2$ ground doublet. Fitting the susceptibility in the LT range to the Curie-Weiss law 
\begin{equation*}
\chi =  \frac{C}{T-\Theta_{\rm CW}},
\end{equation*}
yields the (nominal) Curie-Weiss temperature $\Theta_{\rm CW} = -9.5(9)$ K and the Curie constant $C = 1.14(5)$.
To get some idea of the exchange scale, we consider a nearest-neighbor Heisenberg model on the pyrochlore lattice where
the implied exchange $J$ would be given by
\begin{equation*}
J/{k_B} =  - \frac{3 \Theta_{\rm CW}}{z S(S+1)}\sim 7 \K,
\end{equation*}
here $k_B$ is the Boltzmann constant, $z = 6$ is the coordination number and $S=1/2$. These values are consistent with those reported in Refs. \onlinecite{Reotier2017, Higo2017}.

From the temperature dependence of magnetic susceptibility we can also extract the effective magnetic moment
\begin{equation*}
\mu^2_{\rm eff}={3 k_B/(N_A\mu_B^2)\chi T},
\end{equation*}
here $\mu_B$ is the Bohr magneton and $N_A$ is Avogadro's number. In the LT range  $\mu_{\rm eff}$ is 3.03 $\mu_B/{\rm Yb}^{3+}$, while 
in the HT regime $\mu_{\rm eff}$ saturates at $4.5 \mu_B/{\rm Yb}^{3+}$ (inset of Fig.~\ref{fig7}), close to the (expected) free-ion value $4.54 \mu_B$. 
%{\red {Given the presence of CEF levels at $\sim 30 \meV$, this level of agreement is perfectly satisfactory}.}

\begin{figure}
\centering 
\includegraphics[width=0.5\columnwidth]{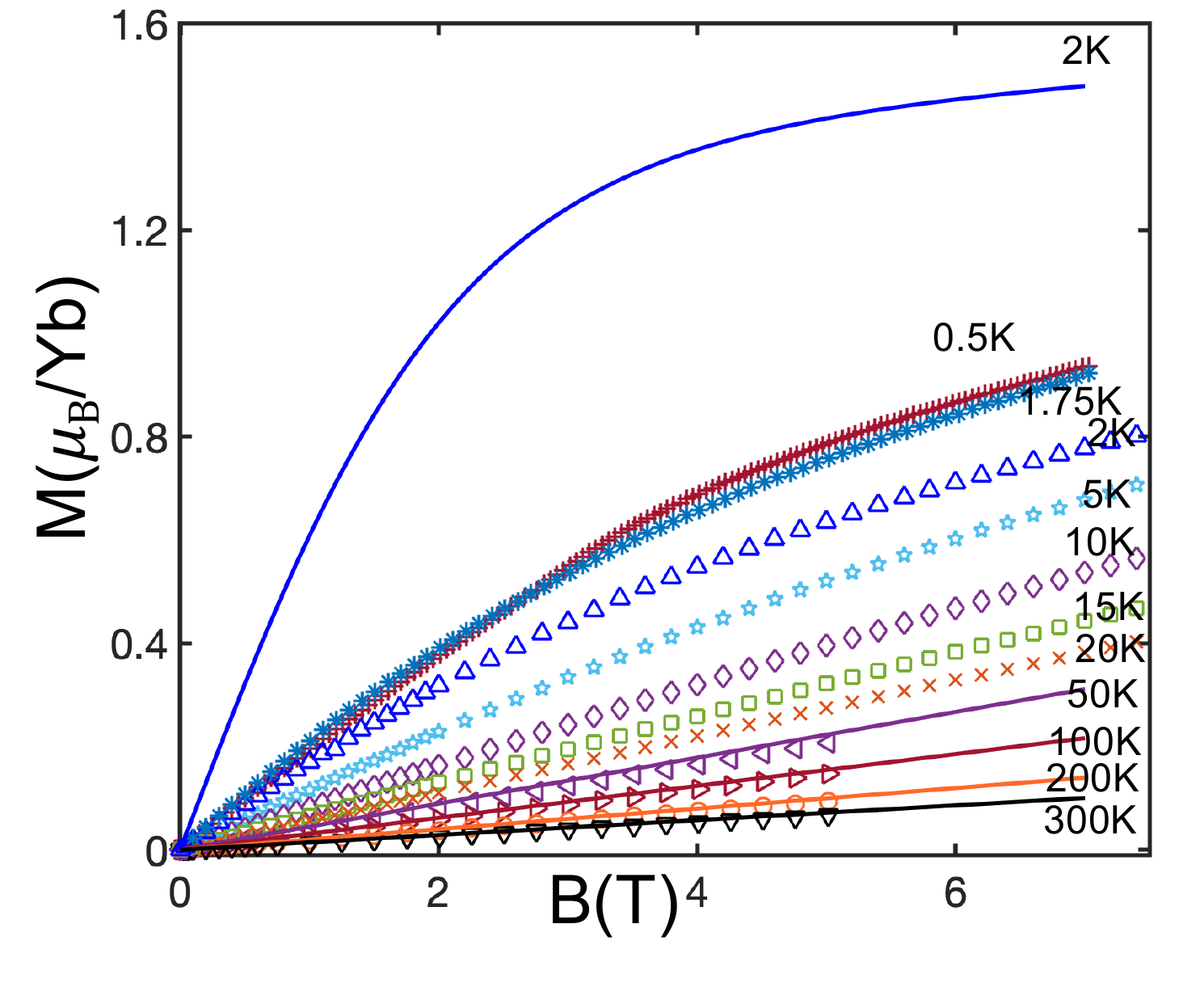} 
\caption{Field dependence of the isothermal magnetization. Experimental data are shown by symbols. The solid lines represent calculated \cite{DucLe} magnetization in the HT regime and at 2 K.}
\label{fig5} 
\end{figure}

Magnetization measured at various temperatures in the 2 K - 300 K range (Fig.~\ref{fig5}) does not saturate in a field of 7 T, reaching only 0.9 $\mu_B$/Yb at 2 K, likely due to the effects of the (relatively large) exchange interactions.

\begin{figure}
\centering 
\includegraphics[width=0.97\textwidth]{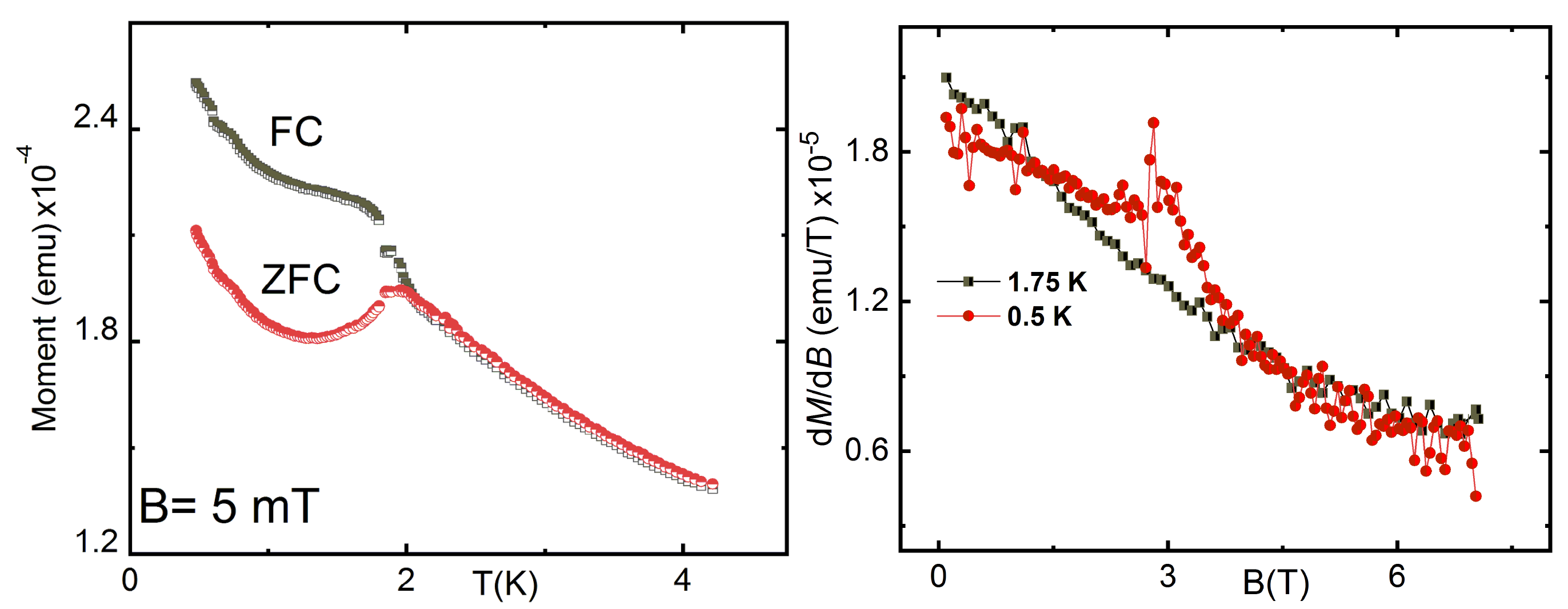} 
\caption{Left: Magnetization for the sample cooled in zero field (ZFC) (red symbols) and in an applied magnetic field (FC) $B$= 5 mT (black symbols). Right: Magnetic field derivative of the isothermal magnetization $dM/dB$ at T=0.5 K (red symbols) and T=1.75 K (black symbols).}
\label{fig1} 
\end{figure}

In the SC regime a pronounced anomaly in susceptibility at $T_N$=1.8 K signals the onset of long-range magnetic order (LRO). This LRO is fragile and the susceptibility $\chi =M/B$ differs below $T_N$ when the sample is cooled in zero-field (ZFC) and in magnetic field (FC), even if the applied field is as small as 5 mT. Fig.~\ref{fig1}, left shows the bifurcation of the susceptibility at $T_N$ for the ZFC and the $B$=5 mT FC states. We estimate that the additional net ferromagnetic (F) component in the FC state is only 1.5$\cdot 10^{-5} \mu_B$/Yb. This bifurcation was observed also by Higo \textit{et al.},\cite{Higo2017} who suggested that the F-signal arises from domain walls formed below $T_N$. A similar FC/ZFC difference was also observed in  Er$_2$Ti$_2$O$_7$\cite{ErTiOZFC}.

A significant change of the long-range order takes place at magnetic fields near $B_c$=3 T.
The corresponding anomaly is well pronounced in the derivative d$M$/d$B$ at 0.5 K compared to the monotonous behaviour at 1.75 K (Fig.~\ref{fig1} right). We interpret this observation as a field-induced crossover between different ordered states and use neutron powder diffraction to study the microscopic origin of this anomaly, as presented in Section ~\ref{Sec3Sub4}.

\subsection{Neutron diffraction in zero and applied magnetic fields}{\label{Sec3Sub4}}

\begin{figure}
\centering 
\includegraphics[width=0.97\columnwidth]{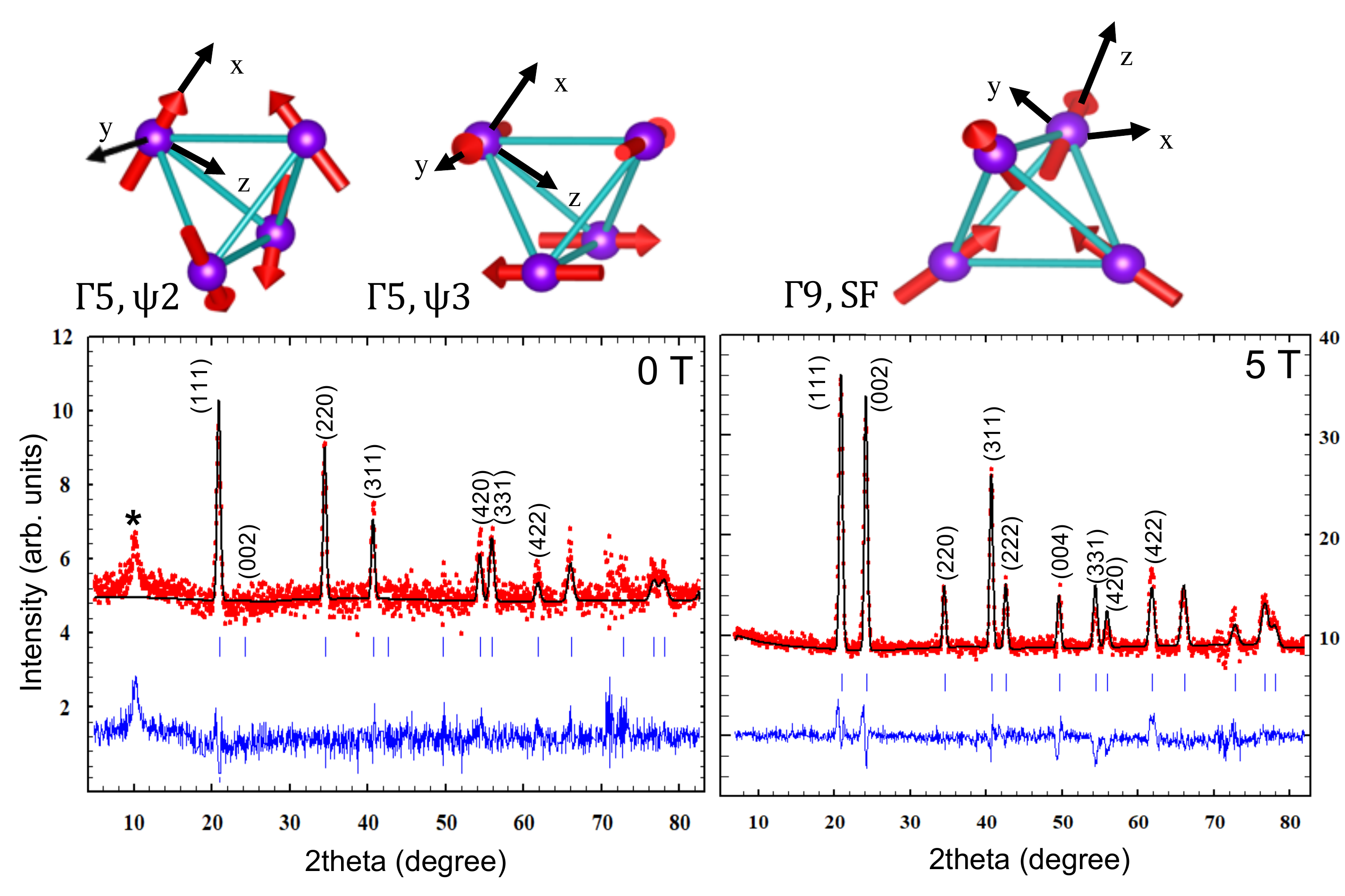} 
\caption{Top: Magnetic moment arrangements for the $\psi_2$ and $\psi_3$ states of $\Gamma_5$ and the splayed $\Gamma_9$ ferromagnet. Bottom:
0.45 K-2.5 K observed, calculated and difference neutron diffraction patterns at 0 T (left) and 5 T (right). The vertical markers designate the $\bf{k}$=0 reflections. The broad peak at 2$\theta$=10 deg is an impurity.\cite{note1}}
\label{fig6} 
\end{figure}

Our zero field neutron powder diffraction patterns (Fig.~\ref{fig6} left) are in agreement with results of Ref. \onlinecite{Reotier2017}.
At the base temperature of 0.45 K we find magnetic intensity at the positions of the ${\bf k}$=0 propagation vector.\cite{note1} No significant difference between the ZFC and 75 mT FC states is detected, which is not surprising as the inferred ferromagnetic component of the FC state seen in the susceptibility is tiny.

\begin{figure}
\centering 
\includegraphics[width=0.49\textwidth]{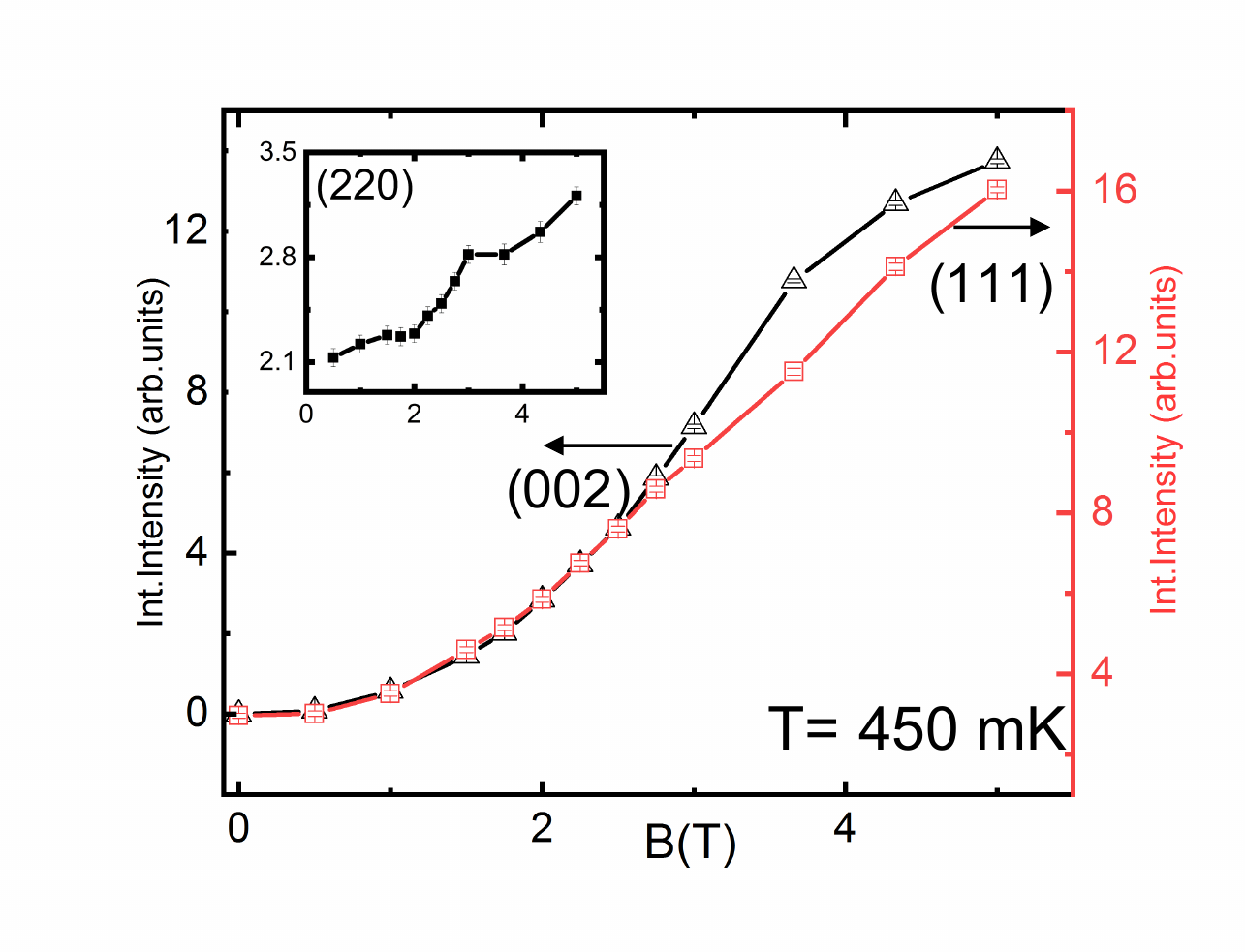} 
\includegraphics[width=0.49\textwidth]{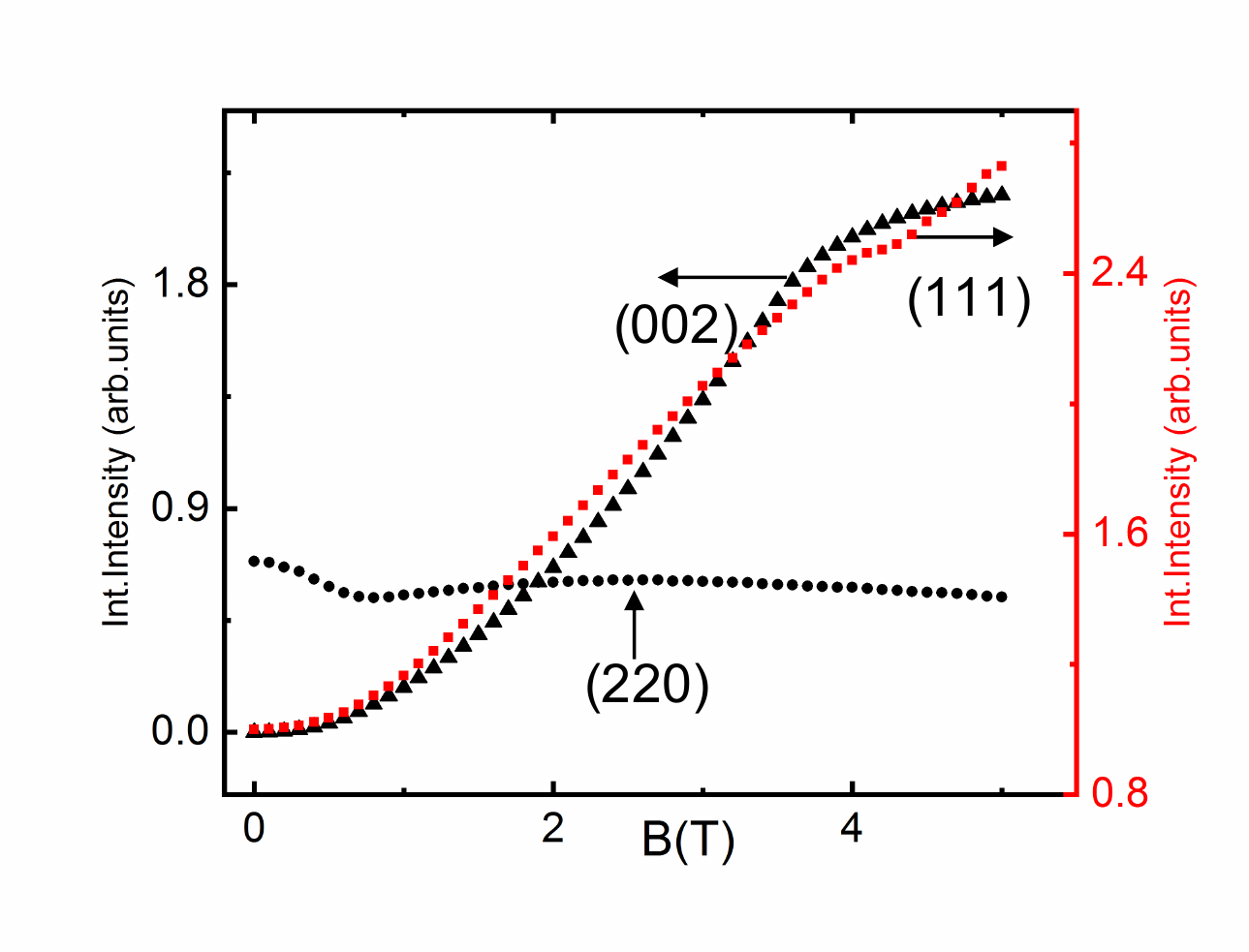} 
\caption{Evolution of the magnetic Bragg peaks (002), (111) and (220) in applied magnetic field in neutron diffraction patterns (left) and according to our theoretical model (right).}
\label{fig3} 
\end{figure}

The more significant changes in diffraction patterns take place at higher fields (Fig.~\ref{fig6} right). The rise of intensity of two main magnetic reflections (0 0 2) and (1 1 1) starts to differ at  $B_c$=3 T (Fig.~\ref{fig3}), where the slope of intensity for the (2 2 0) reflection also changes (Fig.~\ref{fig3} inset). Due to the polycrystalline nature of the sample, we cannot disentangle $B_c$ values for the different crystallographic directions. At 1.2 K the disparity happens at the even lower field of $\sim$2 T (not shown), which implies a tendency of lowering of $B_c$ with temperature increase. 

\subsubsection{Low-energy excitations by neutron spectroscopy}
\begin{figure}%[!]
\centering 
\includegraphics[width=0.97\columnwidth]{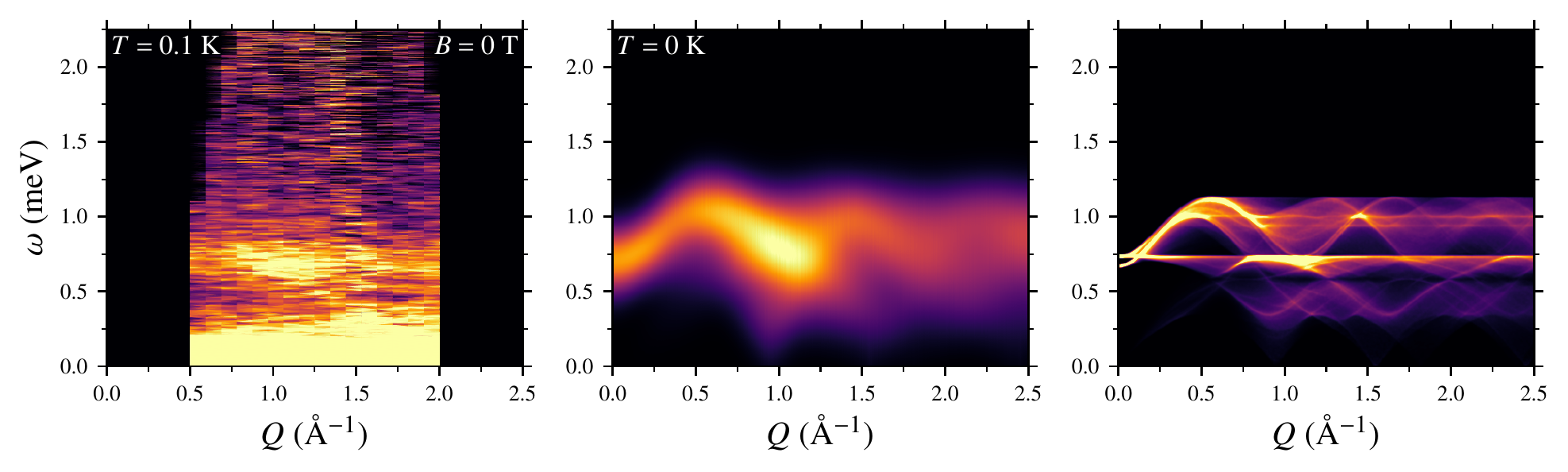}
\includegraphics[width=0.97\columnwidth]{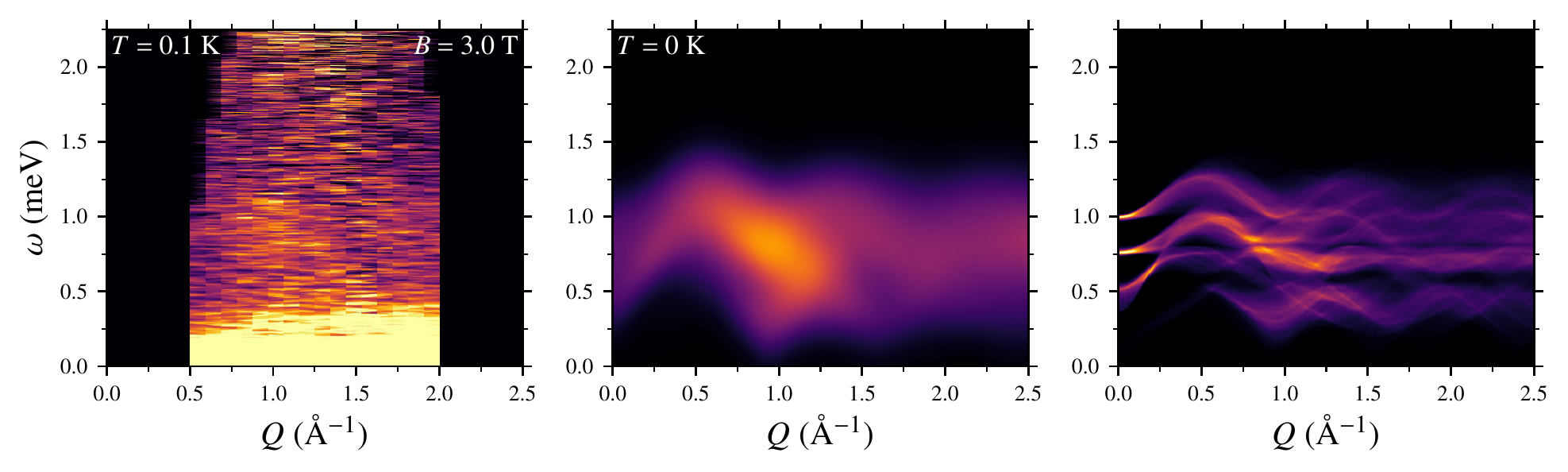}
\includegraphics[width=0.97\columnwidth]{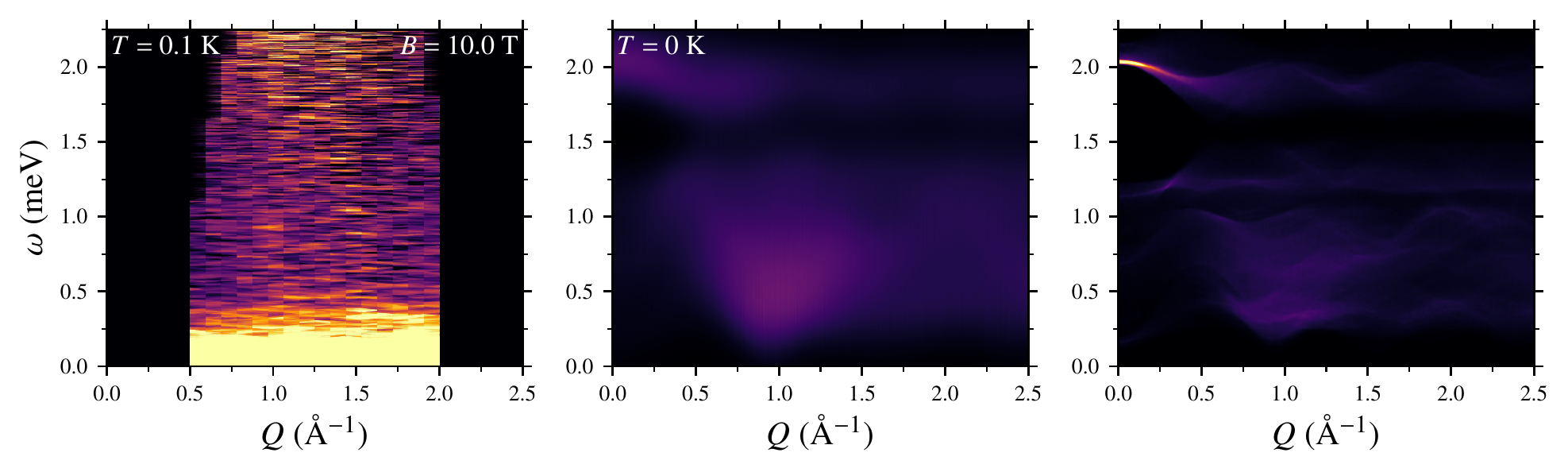}
\caption{Low-energy excitation spectrum of {\CYS} at B= 0 T (top), 3 T (middle), 10 T (bottom). Left:  inelastic neutron scattering measured on FOCUS with $E_i$= 3.7 meV at T= 0.1 K. Right: INS for our model with the exchange parameters of {\CYS} computed within linear spin-wave theory at T= 0 K. The powder average only includes $\vhat{Q}$ $\perp$ $\vhat{B}$ for each randomly sampled $\vhat{B}$, as given in Eq.~(\ref{eq:avg}).
Middle: Calculated spectra from the right panel broadened to better match the experimental resolution.
}
\label{fig4} 
\end{figure}

{\label{Sec3Sub5}}
Fig.~\ref{fig4} (top panel left) presents the low-energy excitations measured with the $E_i$= 3.7 meV setup on FOCUS at the base temperature of 0.1 K. The weakly dispersive band lying around $\hbar \omega$= 0.72 meV and the dispersive branches starting at $\mid\bf{Q}\mid$ = 1.1 \AA$^{-1}$~ vanish at $T_N$ and change with applied magnetic field, thus they are certainly magnetic. These excitations diminish with magnetic field and are not observed for $B >$ 3 T (Fig.~\ref{fig4} (bottom panel left), where the anomaly in magnetization and neutron diffraction is also observed. No excitation could be detected for the setup with higher incoming energy $E_i$= 10.4 meV presumably due to the weakness of the signal.
%
%--------------------------------------------------------
\section{Discussion}{\label{Sec4}}
%--------------------------------------------------------
We first present the crystal-electric field scheme of Yb$^{3+}$ based on our experimental findings and then discuss the model of cooperative phenomena and its agreement with the experimental results.
\subsection{Crystal-electric field parameters}{\label{Sec4Sub1}}
To characterize the crystal-electric field scheme of Yb$^{3+}$ we used the CEF Hamiltonian:
\begin{equation}
H= B_0^2 O_0^2 + B_0^4 O_0^4 + B_3^4 (O_3^4 - O_{-3}^4) + B_0^6 O_0^6 + B_3^6 (O_3^6 - O_{-3}^6) + B_6^6 (O_6^6 - O_{-6}^6),
\label{eq:eq1}
\end{equation}
 with the local quantization axis $<$111$>$, Stevens operators  $O_n^m$ and the corresponding coefficients $B_n^m$.\\
\begin{table}
\caption{The crystal electric field parameters for {\CYS}. The columns list  the experimental (INS) $E_{obs}$ and the calculated $E_{calc}$ energies, the $\mid J_z>$ components of the wave functions of the CEF levels, their $g_{calc}$ and magnetic moment $m_{calc}$ values. For comparison the ground state of CdYb$_2$S$_4$ (Ref.~\onlinecite{Higo2017}) is presented in the second row.\\
\label{tab1}}
\begin{ruledtabular}
\begin{tabular}{cccccccccc}
$E_{obs}$ (meV)&$E_{calc}$ (meV)&$\mid \pm \frac{7}{2}>$&$\mid \mp \frac{5}{2}>$&$\mid \pm \frac{3}{2}>$&$\mid \pm \frac{1}{2}>$&\multicolumn{2}{c}{$g_{calc \parallel, \perp}$}&$m_{calc}$ ($\mu_B$)&\\
\hline
0&0&-0.268& 0.882&0& $\pm$0.388&3.67&2.11&1.37&\\
& 0&-0.301& 0.804&0& 0.513&\multicolumn{2}{c}{2.67}&1.33&[\onlinecite{Higo2017}]\\
28.33(2)&28.34&0.848& 0.407&0& $\mp$0.339&4.93&1.56& 1.55&\\
31.07(2)&31.37&0& 0&1& 0&3.42&0&0.99&\\
&63(3)&$\mp$0.457&$\mp$0.239& 0&-0.857&2.18&4.00&1.75&\\
\end{tabular}
\end{ruledtabular}
\end{table}
The positions and intensities of the CEF excitations observed in the Merlin INS experiment were used to refine the CEF parameters with the simulated annealing algorithm of McPhase \cite{Rotter2004}. 
The parameter space which fits the data is rather wide. Incorporating susceptibility or magnetization data does not help to converge to a unique solution as only the HT regime can be used to fit the CEF parameters.
To confine the fitting parameter space we decided to start simulated annealing from the CEF parameters estimated by Higo \textit{et al.}\cite{Higo2017} from susceptibility data for CdYb$_2$S$_4$. 
This resulted in the following
set of Stevens coefficients: $B_0^2$= -0.397, $B_0^4$= 0.026, $B_3^4$= 0.531, $B_0^6$= 0.0001, $B_3^6$= -0.005, $B_6^6$= 0.003 meV. 
Table~\ref{tab1} presents the resulting calculated energies and composition of the Kramers doublets, as well as their magnetic parameters.

The predicted wave function of the ground state is composed of the dominant $J_z=\pm \frac{5}{2}$ component and significant $J_z=\pm \frac{1}{2}$ and $J_z=\pm \frac{7}{2}$ contributions. It is weakly Ising with the effective $g$-factor components  $g_\parallel$= 3.67,  $g_\perp$= 2.11 with respect to the local (111) axis. The calculated magnetic moment of the ground state is, thus  $m=\frac{1}{2} \mu_B \sqrt \frac{2g_\perp^2 + g_\parallel^2}{3}$=1.37 $\mu_B$/Yb.
The calculated energies of the two first excited levels fits the two observed INS excitations reasonably well. The highest doublet should be located at 63 meV according to our model, however due to a
vanishingly small INS matrix element for the transition between the lowest and the highest doublets the intensity of this excitation is negligibly small.\\
Magnetization calculated for our CEF scheme using the SAFiCF code \cite{DucLe} gives good agreement to magnetization measured above the LT regime (Fig.~\ref{fig5}). In the LT regime the single-ion model is no longer valid due to exchange-induced correlations and agreement is lost.
%
%--------------------------------------------------------
\subsection{Cooperative phenomena}{\label{Sec4Sub2}}
%
%--------------------------------------------------------
\subsubsection{Magnetically ordered states from neutron diffraction}{\label{Sec4Sub2Sub1}}
When determining the magnetic ground state and its evolution with applied magnetic field from neutron powder diffraction we followed the notations and symmetry analysis of Dalmas de Reotier \textit{et al.} \cite{Reotier2017}
The distribution of magnetic intensity in our D20 experiment is identical to the one reported in Ref.~[\onlinecite{Reotier2017}] and our refinements confirm that the $\bf{k}$=0 magnetic order is described by the $\Gamma_5$ irreducible representation (IRR) with the basis vectors (BV) $\psi_2$ or $\psi_3$. The magnetic moments are confined to planes perpendicular to the local $\braket{111}$ axes of the tetrahedron for both BVs, being coplanar for $\psi_3$ (along $\braket{110}$ axes) and noncoplanar for $\psi_2$ (along $\braket{11\bar{2}}$ axes).
These arrangements are indistinguishable by NPD and both result in 0.634(5) $\mu_B$/Yb ordered moment at 0.45 K in zero field, which in close to 0.77(1) $\mu_B$/Yb reported by Dalmas de Reotier \textit{et al.} \cite{Reotier2017} This value is significantly lower than the single-ion expectation value of 1.37 $\mu_B$ for the GS doublet and $\mu_{\rm eff}$ in the LT regime extracted from susceptibility suggesting strong quantum fluctuations.

To understand the microscopic picture at higher fields we performed Rietveld refinements of difference patterns between data measured at a given magnetic field at 0.45 K and data measured at zero field at 2.5 K. Firstly the isotropic powder averaging was considered. Already at $B$=0.5 T we can fit the pattern by a mixture of the $\Gamma_5$ $\psi_2$/$\psi_3$ and the $\Gamma_9$ IRRs with the basis vectors $\psi_7$ and $\psi_8$, which suggests a crossover between these two states. While the  $\psi_8$ BV corresponds to the ferromagnetic (F) component along the $\braket{001}$ axis, the $\psi_7$ BV corresponds to the antiferromagnetic (AF) arrangement along the $\braket{110}$ axes. 
This is the so-called ice-like- splayed ferromagnet (SF) observed %in Yb$_{2}$Sn$_{2}$O$_{7}$ pyrochlore,\cite{Yaouanc2013} which differs from the ice-like SF 
with the $\braket{111}$ local axes reported for Tb$_{2}$Sn$_{2}$O$_{7}$.\cite{Mirebeau2005}
With increasing field the $\Gamma_5$ contribution diminishes and completely vanishes at 3 T, while both $\Gamma_9$ components increase with field.

The fits get worse above 3 T because magnetic field aligns the ferromagnetic $\Gamma_9$ component and isotropic powder averaging cannot be applied anymore. We approximated this grain alignment by the preferred orientation model using a modified March's function ((G cos~$\alpha$)$^2$+$\frac{\textrm{sin}^2 \alpha}{\textrm G})^{-3/2}$. Here G=$T_\perp/T_\parallel$ is a ratio of texture coefficients perpendicular and parallel to the texture vector $\bf{t}$ chosen along the F-component $<$001$>$, $\alpha$ is the acute angle between the scattering vector $\bf Q$ and $\bf{t}$. This way we improve the fits significantly. The obtained values are sensible and the $\psi_7$ and $\psi_8$ BVs of $\Gamma_9$ match the observed intensities. Still this model needs further corroboration by single crystal neutron diffraction.

For our refined model in the maximum applied field of 5 T the F $\psi_8$ component is 0.94(1) $\mu_B$/Yb, which is comparable to the net magnetization 0.796 $\mu_B$/Yb measured for the bulk sample at 0.5 K  and 5 T. Summing the F $\psi_8$ (0.796 $\mu_B$/Yb) and the AF $\psi_7$ (1.26(6) $\mu_B$/Yb) components gives the total ordered moment of 1.58(8) $\mu_B$/Yb, which is comparable to the expected moment 1.37 $\mu_B$ of the GS doublet. The obtained value suggests that the fluctuating part of the moment is suppressed as the magnetic field becomes large, as expected.

%%%%%%%%%%%%%%%%

\begin{table} [h]
\caption{
Magnetic arrangement on the neiboughring Yb-sites 1-4 with the coordinates $xyz$ forming a tetrahedron. From powder neutron diffraction at 0.45 K the moment components are $|M_x|$=$|M_y|$=0.634(5) $\mu_B$/Yb for B=0 T and $|M_x|$=$|M_y|$=0.900(6) $\mu_B$/Yb, $M_z$=0.91(1) $\mu_B$/Yb for B=5 T.
\label{tab2}
}
\begin{ruledtabular}
\begin{tabular}{cccc|ccc}
Site&$x y z$&\multicolumn{2}{c}{B=0 T $\psi_2/\psi_3$ of $\Gamma_5$}&\multicolumn{3}{c}{B=5 T  $\Gamma_9$}\\ 
&&$\psi_2 M_x$&$\psi_2 M_y$&$\psi_7 M_x$&$\psi_7 M_y$&$\psi_8 M_z$\\ 
\hline
1&{\half} {\half} {\half}&+&-&+&+&+\\ 
2&{\half} {\quat} {\quat}&+&+&-&+&+\\ 
3&{\quat} {\half} {\quat}&-&-&+&-&+\\ 
4&{\quat} {\quat} {\half}&-&+&-&-&+\\ 
\end{tabular}
\end{ruledtabular}
\end{table}
%

%--------------------------------------------------------
\subsubsection{Exchange model and comparison to the experiment}{\label{Sec4Sub2Sub2}}
To model {\CYS}, recall the nearest-neighbour anisotropic exchange model for effective $S=1/2$ Kramers doublets on the pyrochlore lattice~\cite{Curnoe2008, Ross2011}. In terms of the pseudo-spin quantized in the \emph{global} frame this can be written as~\cite{Ross2011,Rau2019}
\begin{equation}
    H = \sum_{\avg{ij}} \trp{{\vec{S}}}_i {\mat{J}}_{ij} {\vec{S}}_j - \mu_B \vec{B} \cdot \sum_i \mat{g}_i \vec{S}_i
\label{q:model}
\end{equation}
where ${\mat{J}}_{ab}$ denotes the exchange matrix between
sublattices $a$ and $b$ defined as
\begin{align}
    {\mat{J}}_{12} &= \left(
        \begin{array}{ccc}
             J+K &+\frac{D}{\sqrt{2}} &+\frac{D}{\sqrt{2}}  \\
             -\frac{D}{\sqrt{2}} & J & \Gamma \\
             -\frac{D}{\sqrt{2}} & \Gamma & J
        \end{array}
    \right), & 
    {\mat{J}}_{13} &= \left(
        \begin{array}{ccc}
             J & -\frac{D}{\sqrt{2}} & \Gamma  \\
            +\frac{D}{\sqrt{2}} & J+K &+\frac{D}{\sqrt{2}} \\
             \Gamma & -\frac{D}{\sqrt{2}} & J
        \end{array}
    \right), \nonumber \\    
    {\mat{J}}_{14} &= \left(
        \begin{array}{ccc}
             J & \Gamma & -\frac{D}{\sqrt{2}}  \\
             \Gamma & J & -\frac{D}{\sqrt{2}} \\
            +\frac{D}{\sqrt{2}} &+\frac{D}{\sqrt{2}} & J+K
        \end{array}
    \right), & 
    {\mat{J}}_{23} &= \left(
        \begin{array}{ccc}
             J & -\Gamma &+\frac{D}{\sqrt{2}}  \\
             -\Gamma & J & -\frac{D}{\sqrt{2}} \\
             -\frac{D}{\sqrt{2}} &+\frac{D}{\sqrt{2}} & J+K
        \end{array}
    \right), \nonumber \\    
    {\mat{J}}_{24} &= \left(
        \begin{array}{ccc}
             J &+\frac{D}{\sqrt{2}} & -\Gamma  \\
             -\frac{D}{\sqrt{2}} & J+K &+\frac{D}{\sqrt{2}} \\
             -\Gamma & -\frac{D}{\sqrt{2}} & J
        \end{array}
    \right), & 
    {\mat{J}}_{34} &= \left(
        \begin{array}{ccc}
             J+K & -\frac{D}{\sqrt{2}} &+\frac{D}{\sqrt{2}}  \\
            +\frac{D}{\sqrt{2}} & J & -\Gamma \\
             -\frac{D}{\sqrt{2}} & -\Gamma & J
        \end{array}
    \right).\nonumber
\end{align}
The $g$-factor matrices, $\mat{g}_i$, are given as
\begin{equation}
\mat{g}_i = g_{\pm} \left(\vhat{x}^{}_i\trp{\vhat{x}}_i+\vhat{y}^{}_i\trp{\vhat{y}}_i\right) + g_z \vhat{z}^{}_i\trp{\vhat{z}}_i,
\end{equation}
where $(\vhat{x}_i,\vhat{y}_i,\vhat{z}_i)$ define the local coordinate axes~\cite{Ross2011,Rau2019}. In this basis there are four exchanges: the isotropic Heisenberg exchange ($J$), the Kitaev exchange ($K$), the symmetric off-diagonal exchange ($\Gamma$) and the anti-symmetric Dzyaloshinskii-Moriya (DM) exchange ($D$).

Using theoretical framework of Ref.~\onlinecite{Rau2018} we expect the exchange parameters of {\CYS} to lie in the
regime of dominant antiferromagnetic Heisenberg exchange $J>0$, with subdominant indirect DM, $D<0$, and small symmetric
exchanges $K$ and $\Gamma$ -- that is we expect that $J \gtrsim |D| \gg K,\Gamma$. This scenario puts {\CYS} close to the
phase boundary between the $\Gamma_5$ states and the nearby (splayed) ferromagnet phase. Since the $g$-factors and CEF ground doublet composition from INS data for \CYS{} are not unambiguous, we cannot perform reliable super-exchange calculations, as has been done in Ref.~\onlinecite{Rau2018}. We do note however, that using the $g$-factors presented in Sec. \ref{Sec4}, we find that the values from super-exchange calculations lie precisely in the regime described above.

To determine the exchanges we thus turn to the experimental data. From the Curie-Weiss temperature we can infer the dominant Heisenberg exchange is roughly $J \sim 7$ K. Since a $\Gamma_5$ ground state is found experimentally, we assume that $(K+\Gamma) <0$ and thus such a state is chosen at the classical level. To determine the subdominant (indirect) DM interaction, we consider its effect on the excitation spectrum of the $\Gamma_5$ ground states. We find that the position of the intensity maximum (as seen experimentally) corresponds to a flat band that is directly tuned by the strength of the DM interaction. From this observation, we find that a DM interaction of $D/J \sim -0.3$ best reproduces the zero-field experimental inelastic spectrum (the precise values of $K$ and $\Gamma$ do not strongly affect this conclusion).

Determining the values of $K$ and $\Gamma$ is more difficult. We have fixed $K=\Gamma$ and tuned this common value to capture 
qualitatively the field dependence of the magnetic Bragg intensities; specifically the matching of the field dependence of the upturns in the $[200]$ and $[111]$ intensities. From these considerations we choose $K/J = \Gamma/J = -0.07$, not too far from what has been found via theoretical calculations~\cite{Rau2019}, as well as experimentally in a related breathing pyrochlore material~\cite{Rau2016b}. Our model is thus
\begin{align}
 \label{eq:exch}
 K/J &= -0.07, & \Gamma/J &= -0.07, & D/J &= -0.3,
\end{align}
where $J \sim 7\K$.
At zero field the model results in the $\Gamma_5$ ground state, as is obtained in the experiment.
Like in Yb$_2$Ti$_2$O$_7$, these exchanges are close to the phase boundary
between the SF and $\Gamma_5$ phases, suggesting an important role of the competition between these phases in {\CYS}.
Linear spin-wave theory (LSWT) predicts that quantum fluctuations would
select the $\psi_3$ state out of this manifold through
order-by-quantum-disorder~\cite{Zhito2012,Savary2012}, similar to what has been
suggested for Yb$_2$Ge$_2$O$_7$.

To compare the model and experimental results in an applied magnetic field (both diffraction and inelastic neutron scattering)
we emulate the powder averaging, as well the particular neutron scattering geometry, where
the collected wave-vectors lie roughly in the plane perpendicular to
the field. This results in a correlated average over field
orientation and wave-vector $\sim \int d\vhat{B} \int_{\vhat{Q} \perp
\vhat{B}} d\vhat{Q}$ for both quantities.

To be explicit, for the magnetic Bragg peaks we consider
\begin{equation}
I(\vec{Q};\vec{B}) \propto F(\vec{Q})^2 \sum_{\mu\nu} \left(\delta_{\mu\nu}- \hat{Q}_\mu\hat{Q}_\nu\right) \cc{{M}^{\mu}_{\vec{Q}}(\vec{B})}{M}^{\nu}_{\vec{Q}}(\vec{B})
\end{equation}
where $\vec{M}_{\vec{Q}}(\vec{B}) \equiv \sum_{\vec{r}} e^{i\vec{Q}\cdot\vec{r}} \vec{M}_{\vec{r}}(\vec{B})$ are the
Fourier transforms of the classical ground states for a given field $\vec{B}$ and $F(\vec{Q})$ is the Yb$^{3+}$ magnetic form factor in the dipole approximation. To compare intensities of different peaks in powder patterns the Lorentz factor applied.

For the inelastic spectrum we compute
\begin{equation}
I(\vec{Q},\omega;\vec{B}) \propto F(\vec{Q})^2\sum_{\mu\nu} \left(\delta_{\mu\nu}- \hat{Q}_\mu\hat{Q}_\nu\right) S_{\mu\nu}(\vec{Q},\omega;\vec{B}),
\end{equation}
where $S_{\mu\nu}(\vec{Q},\omega;\vec{B})$ is the dynamical structure factor in a magnetic field $\vec{B}$.
We compute this quantity within LSWT, optimizing for a new classical ground state (assuming the expected $\bf{k}$=0 structure) for each field $\vec{B}$ considered (direction and magnitude).
Accounting for the powder averaging gives
\begin{equation}
 \label{eq:avg}
 I(Q,\omega;B) \propto \int d\vhat{B} \int_{\vhat{Q} \perp \vhat{B}} d\vhat{Q}\ I(Q\vhat{Q},\omega;B\vhat{B}),
\end{equation}
with a similar expression for the Bragg peak intensity
\begin{equation}
I_0(Q;B) \propto \int d\vhat{B}  \int_{\vhat{Q} \perp \vhat{B}} d\vhat{Q}\ I_0(Q\vhat{Q};B\vhat{B}).
\end{equation}
At zero-field we consider the expected $\psi_3$ ground state, averaging over its six domains, though we note that it is difficult to distinguish $\psi_3$ from the $\psi_2$ states or a simple average over the $\Gamma_5$ manifold.

We show the associated evolution of the magnetic Bragg peaks in Fig.~\ref{fig3} (right). The qualitative features
of the Bragg evolution are captured by the theoretical calculation. In particular, one can identify a crossover near $2-3$ T where the correlations change from antiferromagnetic to ferromagnetic. 
The evolution of these peaks is qualitatively different from what is observed in other $\Gamma_5$ ordered magnets, such as Er$_2$Ti$_2$O$_7$; this is a consequence of \emph{Ising}-like $g$-factors of the GS doublet in {\TYX} ($g_z>g_{\pm}$), compared to the $XY$-like $g$-factors ($g_z < g_{\pm}$) found in Er$_2$Ti$_2$O$_7$. Some details, such as the range of the crossover field or evolution of relative intensity of the peaks, are sensitive to the precise values of the $g$-factors, and the small symmetric anisotropies $K$ and $\Gamma$. Since these quantities are not presently known precisely, a quantitative agreement between theory and experiment should not be expected.

We note that the agreement of the absolute ratios of the intensities at low-fields is
not captured by the simple classical calculations. This is likely related to the strong
reduction of the ordered moment observed at zero-field, similar to what has been reported
in the Yb$_2$B$_2$O$_7$ compounds (B = Ti, Sn and Ge). This reduction could be related
to quantum fluctuations and frustration and should thus be relieved as the magnetic
field is increased, as is observed. When determining
the ratios $K/J = \Gamma/J = -0.07$ we thus focused on the high-field values and the field dependence of the intensities, ignoring inconsistencies in the overall relative intensities that
appear at low field (vs. high-field).

We discuss now the inelastic spectrum shown in Figs.~\ref{fig4}.  To
facilitate comparison of the calculated spectra with the experimental
data, we have included a Gaussian broadening to emulate the finite
experimental resolution (middle panel).  The spectra contain a broad
maximum near $Q \sim 1$ \AA~ and $\omega \sim 0.7 \meV$ (used to fix
$D/J = -0.3$) which matches well the experimental spectra presented in
Fig.~\ref{fig4} (left panel).  Note that the overall bandwidth of these
excitations at zero-field is set by the overall scale $J$, giving a
cross-check on the exchange scale determined from the Curie-Weiss fit.

The sharper features present in the INS spectrum at zero field persist to fields of order $2 - 3$ T and then start to dissipate. At larger fields the spin-waves quickly lose intensity as they spread out over a large energy range. This is qualitatively consistent with the experimental data. It can be understood as a consequence of the magnetic anisotropy of this system, in both the single-ion properties and the exchange interactions, causing the excitation spectrum to depend strongly on the field direction. The powder averaging thus produces a variety of quite different spectra that, when averaged, become diffuse without any distinctive feature in the range of wave-vectors probed.

\section{Outlook}

Summarising our findings we find that the {\CYS} spinel shows strong similarities to the ytterbium pyrochlore oxides Yb$_2$B$_2$O$_7$ (R= Yb, Er and B = Sn, Ti, Ge). It is worth mentioning that $XY$ pyrochlore magnets have a remarkably rich phase diagram\cite{Wong2013}, with several competing $\bf{k}$=0 magnetic orders, such as the Palmer-Chalker states, the $\Gamma_5$ manifold with noncoplanar $\psi_2$ and collinear $\psi_3$ states, and splayed ice-like and $XY$-like ferromagnets.
The zero field ordered state in {\CYS}, similarly to {\ETO}, {\EGO} and {\YGO}, belongs to the $\Gamma_5$ irreducible representation. In this $\Gamma_5$ manifold the classical degeneracy between these states can be lifted by several different competing order-by-disorder mechanisms, such as thermal-, quantum-, structural- or virtual-crystal-field-fluctuations.\cite{Zhito2012,Savary2012,Maryasin2014,McClarty2014,McClarty2015,Rau2016a,Rau2019} From our powder diffraction data we cannot distinguish between the expected selection of $\psi_2$ or $\psi_3$ states, though our calculations suggest a $\psi_3$ ground state for {\CYS}.

Identification of the selected state from the $\Gamma_5$ manifold is difficult. In {\ETO} theoretical predictions, based on exchanges determined at high fields, and the experimental ground state agree\cite{Savary2012, Wong2013, Poole2007, Lhotel2017}, with the $\psi_2$ ground state chosen. This selection manifests in single crystal neutron diffraction~\cite{Poole2007} and in the behavior under small magnetic fields,\cite{Lhotel2017}, as well as in more indirect probes such as the opening of a small pseudo-Goldstone gap~\cite{ross2014,petit2014}. Definite identification of the ordered state in {\CYS} requires experiments on single crystals.

More direct parallels can be drawn to the ytterbium pyrochlores. The germanate {\YGO} seems to be the closest analogue to the {\CYS} spinel, though given the limited information available on this compound, we focus our attention on {\YTO},
which has been more extensively studied.\cite{Ross2009, Chang2012, Robert2015, Gaudet2016, Thompson2017} In {\YTO} the ground state was long disputed, due to sample-to-sample variations in experimental properties. The present consensus is that the ground state of {\YTO} is a splayed ice-like ferromagnet. Mysteries still remain however, such as in the diffuse rods of scattering along the $\braket{111}$ directions above $T_C$, and in the absence of well defined spin waves at zero field~\cite{Thompson2017}. The origin of the short-range correlations giving rise to rods can be partly
understood as a result of proximity of {\YTO} to the $\Gamma_5$ phase boundary~\cite{Jaubert2015,Yan2017}, while a complete understanding of the magnetic excitations is still missing. Open questions include the role of strong quantum fluctuations and the importance of extrinsic effects such as disorder play in determining the magnetic spectrum.

Given the {\CYS} spinel appears to be located in the same interesting region of the phase diagram that has driven interest in the Yb$_2$B$_2$O$_7$
pyrochlores, one might hope that, given its different crystal structure, it may shed some light on the outstanding issues in these compounds. Research in pyrochlore magnets distinct from the rare-earth pyrochlore oxides already proved to be very useful in related compounds. For example, in the dipolar spin ice state of the
CdEr$_2$Se$_4$ spinel,\cite{Gao2018} the monopole dynamics appears to be much faster then in the canonical spin ice Dy$_2$Ti$_2$O$_7$ and thus it can bring insight into processes which require equilibration at much lower temperatures that pyrochlore titanates allow.

%--------------------------------------------------------
\section{Conclusions}{\label{Sec5}}
Our experimental findings and theoretical modelling reveal that low-temperature magnetic properties of the {\CYS} spinel are dominated by the well-isolated ytterbium ground state doublet with effective spin $S=1/2$. The CEF analysis, combined with the evolution of the magnetic structure in field, suggests a weak Ising character of the ytterbium ground state doublet. Nevertheless the anisotropic exchange interactions of {\CYS} select a ground state of the $XY$-type, with long-range antiferromagnetic $\Gamma_5$ order below $T_N=1.8$ K. Through comparison of the inelastic spectrum with theoretical models, we find that the properties of {\CYS} are consistent with expectations that it lies close to the phase boundary between the $\Gamma_5$ states and an SF phase, similar to the Yb$_2$B$_2$O$_7$ pyrochlores. Further growth of single crystals of the {\CYS} spinel will allow to explore this physics, and test the universality of some of the more exotic features (such as the broad continuum of excitations) that have been observed in other ytterbium pyrochlores.
%%%%
\newpage
%--------------------------------------------------------
%--------------------------------------------------------
%--------------------------------------------------------
%--------------------------------------------------------
%--------------------------------------------------------
%--------------------------------------------------------
%--------------------------------------------------------
%
%--------------------------------------------------------
\newpage
\begin{acknowledgments}
This work was performed at SINQ, Paul Scherrer Institute, Villigen, Switzerland with financial support of the Swiss National Science Foundation (SNF) (Grant No. 200021-140862). This work was in part supported by Deutsche Forschungsgemeinschaft
(DFG) (Grant SFB 1143) and  SNF (Grant No. 206021-139082). We thank S. Gao, M. Pregelj, M. J. P. Gingras for discussions.
%We thank J. Embs, L. Keller, M. Medarde, S. Tian, A. Cervellino for experimental support and S. Gao, M. Pregelj, M. J. P. Gingras for discussions.
\end{acknowledgments}

 %--------------------------------------------------------

%
\end{document}